\documentclass[showpacs,twocolumn,floatfix]{revtex4}
\usepackage{graphicx}
\usepackage{amsmath}

\input{tcilatex}

\begin{document}

\title{Shot noise spectrum of superradiant entangled excitons}
\author{Y. N. Chen$^{1}$, T. Brandes$^{2}$, C. M. Li$^{1}$, and D. S. Chuu$%
^{1}$}
\affiliation{$^{1}$Department of Electrophysics, National Chiao-Tung University, Hsinchu
300, Taiwan\\
$^{2}$Department of Physics, UMIST, P.O. Box 88, Manchester, M60 1QD, United
Kingdom}
\date{\today}

\begin{abstract}
The shot noise produced by tunneling of electrons and holes into a double
dot system incorporated inside a p-i-n junction is investigated
theoretically. The enhancement of the shot noise is shown to originate from
the entangled electron-hole pair created by superradiance. The analogy to
the superconducting cooper pair box is pointed out. A series of Zeno-like
measurements is shown to destroy the entanglement, except for the case of
maximum entanglement.
\end{abstract}

\maketitle

\section{Introduction}

Quantum entanglement has become one of the most important issues since the
rapid developments in quantum information science \cite{1}. Much research
has %%%
been devoted to studying entanglement as induced by a direct interaction
between the individual subsystems \cite{2}. Very recently, a lot of
attention has been focused on \emph{reservoir-induced} entanglement \cite{3}
with the purpose to shed light on the generation of entanglement, and to
better understand quantum decoherence.

Furthermore, shot noise \cite{4} has been identified as a valuable indicator
of particle entanglement in transport experiments.\cite{5} A well studied
example is the doubling of the full shot noise in \textbf{S-I-N} tunnel
junctions\cite{6}, where \textbf{N} is a normal metal, \textbf{S} stands for
a superconductor, and \textbf{I} is an insulating barrier. The origin of
this enhancement comes from the break-up of the spin-singlet state, which
results in a quick transfer of two electrical charges.

In this paper, we demonstrate how the dynamics of entangled excitons formed
by superradiance can be revealed from the observations of current
fluctuations. A doubled zero-frequency shot noise is found for the case of
zero subradiant decay rate. We relate the particle noise to photon noise by
calculating the first order photon coherence function. Furthermore, strong
reservoir coupling acts like a continuous measurement, which is shown to
suppress the formation of the entanglement, except for the state of maximum
entanglement. These novel features imply that our model provides a new way
to examine both the bunching behavior and a Zeno-like effect of the
reservoir induced entanglement.

\section{Double Dot Model}

The effect appears in double quantum dots embedded inside a \emph{p-i-n}
junction\cite{7}. It involves superradiant\ and subradiant decay through two
singlet and triplet entangled states, $\left| S_{0}\right\rangle =\frac{1}{%
\sqrt{2}}(\left| U_{1}\right\rangle -\left| U_{2}\right\rangle )$ and $%
\left| T_{0}\right\rangle =\frac{1}{\sqrt{2}}(\left| U_{1}\right\rangle
+\left| U_{2}\right\rangle )$, and one ground state $\left| D\right\rangle
=\left| 0,0;0,0\right\rangle $, where $\left| U_{1}\right\rangle =\left|
e,h;0,0\right\rangle $ ( $\left| U_{2}\right\rangle =\left|
0,0;e,h\right\rangle $ ) represents one exciton in dot 1 (2). Electron and
hole reservoirs coupled to both dots have chemical potentials such that
electrons and holes can tunnel into the dot. For the physical phenomena we
are interested in, the current is assumed to be conducted through dot 1 only
(Fig. 1). Therefore, the exciton states $\left| 0,0;e,h\right\rangle $ (in
dot 2) can only be created via the exciton-photon interactions. 

%%%%%%%%%%%%%%%%%%%%%%%%%%%%%%%%%%%%%%%%%%%%%%%%%%%%%%%%%%%%%%%%%%%%
\begin{figure}[th]
\centerline{
    \includegraphics[clip=true,width=0.6\columnwidth]{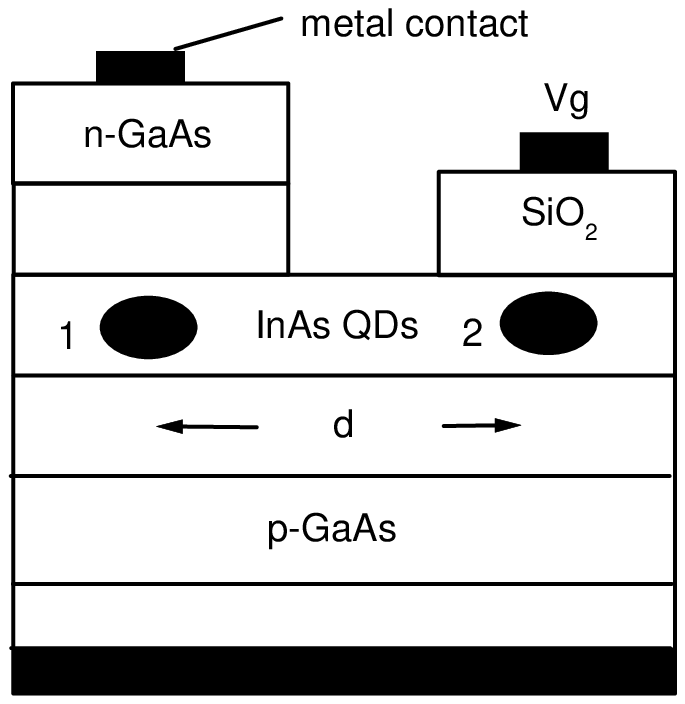}
  }
\caption{Schematic view of the structure.}
\label{disp1}
\end{figure}
%%%%%%%%%%%%%%%%%%%%%%%%%%%%%%%%%%%%%%%%%%%%%%%%%%%%%%%%%%%%%%%%%%%
%%%%%%%%%%%%%%%%%%%%%%%%%%%%%%%%%%%%%%%%%%%%%%%%%%%%%%%%%%%%%%%%%%%%

The exciton-photon coupling is described by an interaction Hamiltonian

\begin{eqnarray}
H_{I} &=&\sum_{\mathbf{k}}\frac{1}{\sqrt{2}}g\{D_{\mathbf{k}}b_{\mathbf{k}%
}[(1+e^{i\mathbf{k}\cdot \mathbf{r}})\left| S_{0}\right\rangle \left\langle
D\right|  \notag \\
&&+(1-e^{i\mathbf{k}\cdot \mathbf{r}})\left| T_{0}\right\rangle \left\langle
D\right| ]+H.c.\},
\end{eqnarray}%
where $b_{\mathbf{k}}$ is the photon operator, $gD_{\mathbf{k}}$ is the
coupling strength, $\mathbf{r}$ is the position vector between the two
quantum dots. Here, $g$ is a constant with the unit of the tunneling rate.
The dipole approximation is not used in our calculation since we keep the
full $e^{i\mathbf{k}\cdot \mathbf{r}}$ terms in the Hamiltonian. The
coupling of the dot states to the electron and hole reservoirs is described
by the standard tunnel Hamiltonian

\begin{equation}
H_{V}=\sum_{\mathbf{q}}(V_{\mathbf{q}}c_{\mathbf{q}}^{\dagger }\left|
0\right\rangle \left\langle U_{1}\right| +W_{\mathbf{q}}d_{\mathbf{q}%
}^{\dagger }\left| 0\right\rangle \left\langle D\right| +H.c.),
\end{equation}%
where $c_{\mathbf{q}}$ and $d_{\mathbf{q}}$ are the electron operators in
the left and right reservoirs, respectively, and $\left| 0\right\rangle
=\left| 0,h;0,0\right\rangle $ denotes one-hole state in dot 1. $V_{\mathbf{q%
}}$ and $W_{\mathbf{q}}$ couple the channels $\mathbf{q}$ of the electron
and the hole reservoirs. Here we have neglected the state $\left|
e,0;0,0\right\rangle $ for convenience. This can be justified by fabricating
a thicker barrier on the electron side so that there is little chance for an
electron to tunnel in advance \cite{8}.

\section{Rate Equations and Noise Spectrum}

The rates $\Gamma _{U}$ (electron reservoir)\ and $\Gamma _{D}$ (hole
reservoir) for tunneling between the dot and the connected reservoirs can be
calculated from $H_{V}$ by perturbation theory. In double quantum dots,
decay of the excited levels is governed by collective behavior, i.e.
superradiance and subradiance. The corresponding decay rates for the state $%
\left| S_{0}\right\rangle $ and $\left| T_{0}\right\rangle $ can be obtained
from $H_{I}$ and are denoted by $g^{2}\gamma _{S}$ and $g^{2}\gamma _{T}$,
respectively. We are then in the position to set up the equations of motion
for the time-dependent occupation probabilities $n_{j}(t),$ $j=$ $0$, $D,$ $%
S_{0},T_{0}$, of the double dot states. Together with the normalization
condition $\sum_{j}n_{j}(t)=1$, the equations of motion are
Laplace-transformed into $z-$space \cite{9} for convenience and read 
\begin{widetext}
\begin{eqnarray}
z \hat{n}_{S_{0}}(z) &=&-ig\left[\hat{p}_{S_{0},D}(z)-\hat{p}_{D,S_{0}}(z)\right]+\Gamma _{U}\left[\frac{1}{z}%
-\hat{n}_{S_{0}}(z)-\hat{n}_{T_{0}}(z)-\hat{n}_{D}(z)\right],  \notag \\
z \hat{n}_{T_{0}}(z) &=&-ig\left[\hat{p}_{T_{0},D}(z)-\hat{p}_{D,T_{0}}(z)\right]+\Gamma _{U}\left[\frac{1}{z}%
-\hat{n}_{S_{0}}(z)-\hat{n}_{T_{0}}(z)-\hat{n}_{D}(z)\right],  \notag \\
z \hat{n}_{D}(z) &=&ig\left[\hat{p}_{S_{0},D}(z)-\hat{p}_{D,S_{0}}(z)+\hat{p}_{T_{0},D}(z)-\hat{p}_{D,T_{0}}(z)
\right]-\frac{2\Gamma
_{D}}{z}\hat{n}_{D}(z).
\end{eqnarray}%
\end{widetext}Here, $p_{S_{0},D}(t)=p_{D,S_{0}}^{\ast }(t)$ and $%
p_{T_{0},D}(t)=p_{D,T_{0}}^{\ast }(t)$ are off-diagonal matrix elements of
the reduced density operator of the double dots, whose Laplace-transformed
equations of motion close the set (3):

\begin{eqnarray}
\hat{p}_{S_{0},D}(z) &=&ig\gamma _{S}\hat{n}_{S_{0}}(z)-\Gamma _{D}\gamma
_{S}\hat{p}_{S_{0},D}(z),  \notag \\
\hat{p}_{T_{0},D}(z) &=&ig\gamma _{T}\hat{n}_{T_{0}}(z)-\Gamma _{D}\gamma
_{T}\hat{p}_{T_{0},D}(z).
\end{eqnarray}

Note that in getting above equations, one has to do a decoupling
approximation of dot operators and photon operators. This means we are
interested in small coupling parameters here, and a decoupling of the
reduced density matrix $\widetilde{\rho }(t^{\prime })$ is used : $%
\widetilde{\rho }(t^{\prime })\approx \rho _{ph}^{0}Tr_{ph}\rho (t^{\prime
}) $. \cite{9} The stationary tunnel current $I$ can be defined as the
change of the occupation of $n_{D}(t)$ for large times $t$ and is given by 
\begin{equation}
I\equiv \lim_{t\to \infty}ig\left[%
p_{S_{0},D}(t)-p_{D,S_{0}}(t)+p_{T_{0},D}(t)-p_{D,T_{0}}(t)\right],
\end{equation}%
where we have set the electron charge $e=1$ for convenience.

In a quantum conductor in nonequilibrium, electronic current noise
originates from the dynamical fluctuations of the current being away from
its average. To study correlations between carriers, we relate the qubit
dynamics with the hole reservoir operators by introducing the degree of
freedom $n$ as the number of holes that have tunneled through the hole-side
barrier and write 
\begin{eqnarray}
\overset{\cdot }{n}_{0}^{(n)}(t) &=&-\Gamma _{U}n_{0}^{(n)}(t)+\Gamma
_{D}n_{D}^{(n-1)}(t),  \notag \\
\overset{\cdot }{n}_{S_{0}}^{(n)}(t) &=&\frac{\Gamma _{U}}{2}%
n_{0}^{(n)}(t)+ig(p_{S_{0},D}^{(n)}(t)-p_{D,S_{0}}^{(n)}(t)),  \notag \\
\overset{\cdot }{n}_{T_{0}}^{(n)}(t) &=&\frac{\Gamma _{U}}{2}%
n_{0}^{(n)}(t)+ig(p_{T_{0},D}^{(n)}(t)-p_{D,T_{0}}^{(n)}(t)),  \notag \\
\overset{\cdot }{n}_{D}^{(n)}(t) &=&-\Gamma
_{D}n_{0}^{(n)}(t)-ig(p_{S_{0},D}^{(n)}(t)-p_{D,S_{0}}^{(n)}(t)  \notag \\
&&+p_{T_{0},D}^{(n)}(t)-p_{D,T_{0}}^{(n)}(t)).
\end{eqnarray}%
Eqs. (6) allow us to calculate the particle current and the noise spectrum
from $%
P_{n}(t)=n_{0}^{(n)}(t)+n_{S_{0}}^{(n)}(t)+n_{T_{0}}^{(n)}(t)+n_{D}^{(n)}(t)$
which gives the total probability of finding $n$ electrons in the collector
by time $t$. In particular, the noise spectrum $S_{I_{D}}$ can be calculated
via the MacDonald formula \cite{10}.

\begin{equation}
S_{I_{D}}(\omega )=2\omega e^{2}\int_{0}^{\infty }dt\sin (\omega t)\frac{d}{%
dt}[\left\langle n^{2}(t)\right\rangle -(t\left\langle I\right\rangle )^{2}],
\end{equation}%
where $\frac{d}{dt}\left\langle n^{2}(t)\right\rangle =\sum_{n}n^{2}\overset{%
\cdot }{P_{n}}(t)$. Solving Eqs. (6) and (3), we obtain

\begin{equation}
S_{I_{D}}(\omega )=2eI\{1+\Gamma _{D}[\hat{n}_{D}(z=-i\omega )+\hat{n}%
_{D}(z=i\omega )]\}.
\end{equation}%
In the zero-frequency limit, Eq. (8) reduces to

\bigskip 
\begin{equation}
S_{I_{D}}(\omega =0)=2eI\{1+2\Gamma _{D}\frac{d}{dz}[z\hat{n}%
_{D}(z)]_{z=0}\},
\end{equation}
which is analogous to a recent calculation of noise in dissipative, open
two-level systems \cite{AB03}.

\section{Results}

\subsection{Current Noise}

To display the dependence of carrier correlations on the dot distance $d$,
Fig. 2 shows the result for zero-frequency noise $S_{I_{D}}(\omega =0)$ as a
function of the inter-dot distance. In plotting the figure, the tunneling
rates, $\Gamma _{U}$ and $\Gamma _{D}$, are assumed to be equal to $%
0.1\gamma _{0}$ and $\gamma _{0},$ respectively. Here, a value of 1/1.3ns
for the free-space quantum dot decay rate $\gamma _{0}$ is used in our
calculations \cite{11}. As shown in Fig. 2, the Fano factor 
\begin{equation*}
F\equiv \frac{S_{I_{D}}(0)}{2e\langle I\rangle }
\end{equation*}%
is enhanced by a factor of 2 as the dot distance $d$ is much smaller than
the wavelength ($\lambda $) of the emitted photon. To explain this
enhancement, we approximate the Fano factor in the limit of small subradiant
decay rate, i.e. 
\begin{equation*}
g^{2}\gamma _{S}\ll \Gamma _{U}<\Gamma _{D}\approx g^{2}\gamma _{T},
\end{equation*}%
where we obtain 
\begin{equation}
\frac{S_{ID}(0)}{2e\langle I\rangle }\approx 2-2g^{2}\gamma _{S}[\frac{1}{%
g^{2}\gamma _{T}}+3(\frac{1}{\Gamma _{D}}+\frac{1}{\Gamma _{U}})+\frac{%
2\Gamma _{D}}{g^{2}}].
\end{equation}%
This is analogous to the case of the single electron transistor near a
Cooper pair resonance as discussed recently by Choi and co-workers \cite{12}%
. In their calculations, the Fano factor is expressed as $\frac{S(0)}{%
2e\langle I\rangle }=2-\frac{8E_{J}^{2}(E_{J}^{2}+2\Gamma ^{2})}{%
(3E_{J}^{2}+\Gamma ^{2}+4\varepsilon ^{2})^{2}}.$ In the strong dephasing
limit ($E_{J}\ll \Gamma $, where $E_{J}$ is the Josephson coupling energy),
the zero-frequency shot noise is also enhanced by a factor of 2.\ Since the
doubled shot noise in Josephson junction is attributed to the bunching
behavior of Cooper pairs (in singlet state), we then conclude the
enhancement in our system is also due to the entanglement induced by the
photon reservoir.

%%%%%%%%%%%%%%%%%%%%%%%%%%%%%%%%%%%%%%%%%%%%%%%%%%%%%%%%%%%%%%%%%%%%
\begin{figure}[th]
\centerline{
    \includegraphics[clip=true,width=0.9\columnwidth]{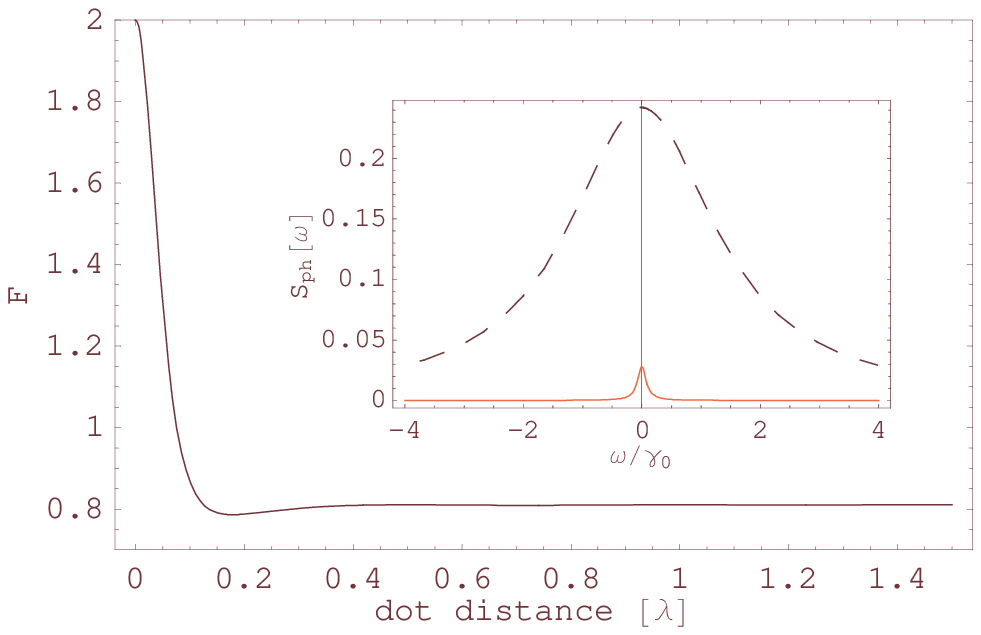}
  }
\caption{Fano factor as a function of inter-dot distance. The vertical and
horizontal units are $\frac{S_{I_{D}}(0)}{2eI}$ and $\protect\lambda $,
respectively. The inset shows the value of $S_{ph}(\protect\omega )$ is
equal to that of one-dot limit for $d\rightarrow \infty $ (dashed line),
while it approaches \emph{zero noise} as $d=0.005\protect\lambda $ (red
line). }
\end{figure}
%%%%%%%%%%%%%%%%%%%%%%%%%%%%%%%%%%%%%%%%%%%%%%%%%%%%%%%%%%%%%%%%%%%
%%%%%%%%%%%%%%%%%%%%%%%%%%%%%%%%%%%%%%%%%%%%%%%%%%%%%%%%%%%%%%%%%%%%

\subsection{Photon Noise}

It is worthwhile to compare the current noise with the \emph{photon noise}
generated by the collective decay of the double dot excitons. In order to do
so, we have calculated the power spectrum of the fluorescence spectrum\cite%
{13}, which can be expressed as

\begin{equation}
S_{ph}(\omega )=\frac{1}{\pi }\func{Re}\int_{0}^{\infty }G^{(1)}[\tau
]e^{i\omega \tau }d\tau ,
\end{equation}%
where $G^{(1)}[\tau ]$ is the first order coherence function and reads

\begin{eqnarray}
G^{(1)}[\tau ] &\propto &\left| 1+e^{i2\pi d/\lambda }\right|
^{2}\left\langle p_{S_{0},D}(0)p_{D,S_{0}}(\tau )\right\rangle  \notag \\
&&+\left| 1-e^{i2\pi d/\lambda }\right| ^{2}\left\langle
p_{T_{0},D}(0)p_{D,T_{0}}(\tau )\right\rangle .
\end{eqnarray}%
The two time-dependent correlation functions in the above equation can be
calculated from the quantum regression theorem, and the numerical result of $%
S_{ph}(\omega )$ is shown explicitly in the inset of Fig. 2. As can be seen,
the value of $S_{ph}(\omega )$ is equal to that of the one-dot case for $%
d\rightarrow \infty $ (dashed line), while it approaches zero as $%
d=0.005\lambda $ (red line). In the limit of $d=0$, one observes no photon
emission from the double dot system since the exciton is now in its maximum
entangled state and does not decay. This feature implies that photon noise
is suppressed by the bunching of excitons, and its behavior is opposite to
that for the electronic case \cite{14}.

\subsection{Noise and Measurement}

Now we investigate how the measurement affects the shot noise spectrum. In
the usual Zeno paradox \cite{15}, a two-level system (qubit) is completely
frozen under a series of measurements, whose time interval $\Delta t$ is
much smaller than the memory time of the reservoir. In our model, the
presence of the exciton state can be viewed as the excited state, and
whether or not the next hole can tunnel in is determined by the occupation
of this state. Similar to the quantum Zeno effect, the tunneling of holes at
the hole-side tunneling rate $\Gamma _{D}$ can be thought as a continuous
measurements. The ''interval time'' $\Delta t$ is then inversely
proportional to $\Gamma _{D}.$ Fig. 3 represents the effects of measurements
on the frequency-dependent shot noise spectrum. The numerical results for
the tunneling rate $\Gamma _{D}=20$ $\gamma _{0}$ and $\Gamma _{D}=$ $\gamma
_{0}$ are demonstrated by solid and dashed curves, respectively. If the
subradiant decay rate is set to zero, one obtains the doubled shot noise as
mentioned above. Without superradiance, the values of the Fano factor are
always below unity as shown by the right inset of Fig. 3. An interesting
feature is that the half-width of the spectrum is narrowed for strong
measurements ($\Gamma _{D}=20$ $\gamma _{0}$). If one increases the
electron-side tunneling rate $\Gamma _{U}$, there exists no such behavior.
This implies that the effective decay rate is reduced in the presence of
strong measurements.

%%%%%%%%%%%%%%%%%%%%%%%%%%%%%%%%%%%%%%%%%%%%%%%%%%%%%%%%%%%%%%%%%%%%
\begin{figure}[th]
\centerline{
    \includegraphics[clip=true,width=0.9\columnwidth]{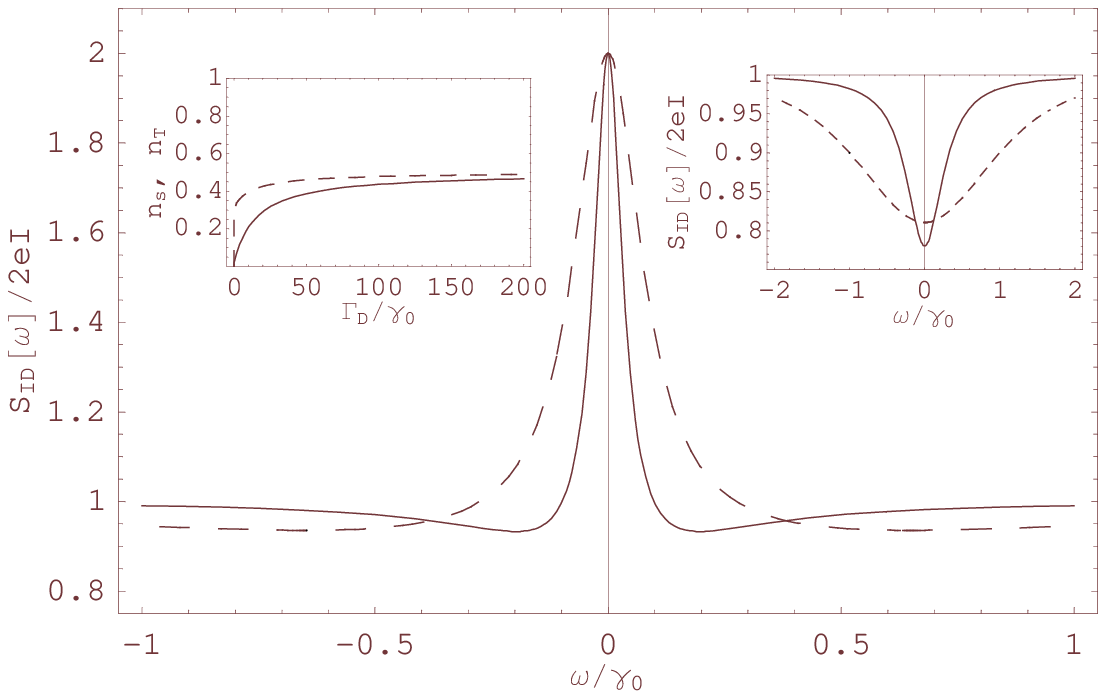}
  }
\caption{Effect of measurements on current noise \ $S_{ID}(\protect\omega )$
( ''maximum'' superradiance, $g^{2}\protect\gamma _{T}=2g^{2}\protect\gamma %
_{0}$ , $g^{2}\protect\gamma _{S}=0$). Solid and dashed lines correspond to $%
\Gamma _{D}=20$ $\protect\gamma _{0}$ and $\Gamma _{D}=$ $\protect\gamma _{0}
$ , respectively. Right inset : the case of no superradiance. Left inset :
expectation value of the excited states $\left\langle n_{S}\right\rangle $
and $\left\langle n_{T}\right\rangle $ as a function of $\Gamma _{D}.$}
\end{figure}
%%%%%%%%%%%%%%%%%%%%%%%%%%%%%%%%%%%%%%%%%%%%%%%%%%%%%%%%%%%%%%%%%%%
%%%%%%%%%%%%%%%%%%%%%%%%%%%%%%%%%%%%%%%%%%%%%%%%%%%%%%%%%%%%%%%%%%%%

To investigate thoroughly the underlying physics, we plot the expectation
value of the excited states $\left\langle n_{S}\right\rangle $ and $%
\left\langle n_{T}\right\rangle $ as a function of $\Gamma _{D}$ in the left
inset of Fig. 3. One clearly finds the occupation probabilities grow with
increasing $\Gamma _{D}$, and both of them approaches the value of $\frac{1}{%
2}$. This not only means the measurements tend to localize the exciton in
its excited state \cite{16}, but also tells us the entanglement is destroyed
under the strong measurements. However, in the limit of no subradiance ($%
g^{2}\gamma _{S}=0$), the occupation probability of the singlet state is
always equal to one, i.e. maximum entanglement is robust against strong
measurements. This is because once the maximum entangled state is formed,
the total probability in the excited states is also maximum. Strong
measurements on the ground state $\left| D\right\rangle $ have no influence
on the singlet entangled state.

\section{Discussion and Conclusion}

A few remarks about experimental realizations of the present model should be
mentioned here. One should note that biexciton and charged-exciton effects
are not included in our present model. Inclusion of these additional states
is expected to suppress the enhancement of the shot noise, i.e. degrees of
entanglement. However, this can be controlled well by limiting the value of
bias voltage so that only the ground-state exciton is present \cite{17}. To
produce the maximum entangled state, one can also incorporate the device
inside a microcavity \cite{18}. There are two advantages of this design: The
maximum entanglement can be generated even for remote separation of the two
dots, and Forster process \cite{19} is avoided at this distance.

As for the problem of decoherence due to interactions with phonons, recent
experimental data have shown that the exciton-phonon dephasing rate is
smaller than the radiative decay one in a quantum dot. This means that due
to the discrete energy level scheme in a quantum dot, the effect of \emph{%
phonon-bottleneck} tends to suppress the exciton-phonon interaction \cite{20}%
. Although the present model describes tunneling of electrons and holes into
semiconductor quantum dots, the whole theory can be applied to electron
tunneling through coupled quantum dots which are interacting via a common
phonon environment \cite{21}.

In conclusion, we have demonstrated that the shot noise of superradiant
entangled excitons is enhanced by a factor of two as compared to the
poissonian value. This enhancement was attributed to exciton entanglement,
induced by the electromagnetic field (common photon reservoir), and an
analogy to the Cooper pair box was made. Second, we found the relaxation
behavior of the qubits in the presence of strong measurements, and the
Zeno-like effect tends to destroy the entanglement and localize the qubits
in the excited states.

This work is supported partially by the National Science Council, Taiwan
under the grant number NSC 92-2120-M-009-010.


\begin{thebibliography}{99}
\bibitem{8} In the following calculations, the electron-side tunneling rate $%
\Gamma _{U}$ is always assumed to be much smaller than the hole-side
tunneling rate $\Gamma _{D}$.
\end{thebibliography}
\end{document}